\documentclass[prb,aps,showpacs,twocolumn]{revtex4}
\usepackage{dcolumn}
\usepackage{bm}
\usepackage{amsmath}
\usepackage{graphics}
\usepackage{epsfig}

\begin{document}

\title{ Microscopic Two-fluid Theory of Rotational Constants
of the OCS-H$_2$ Complex in $^4$He Droplets }

\author{ Yongkyung Kwon$^{1,2}$}
\author{ K. Birgitta Whaley$^2$}
\affiliation{$^1$Department of Physics,
Konkuk University, Seoul 143-701, Korea}
\affiliation{$^2$Department of Chemistry and Kenneth S. Pitzer Center for
Theoretical Chemistry, University of California, Berkeley, CA 94720,
USA}

\date{\today}

\begin{abstract}

We present a microscopic quantum analysis for rotational constants of the OCS-H$_2$ complex in helium droplets using the local two-fluid theory 
in conjunction with path 
integral Monte Carlo simulations. Rotational constants are derived from effective moments of 
inertia calculated assuming that motion of the H$_2$ molecule and the local non-superfluid 
helium density is rigidly coupled to the molecular rotation of OCS and employing path
integral methods to sample the corresponding H$_2$ and helium densities. 
The rigid coupling assumption for H$_2$-OCS is calibrated by comparison with exact calculations of the free OCS-H$_2$ complex.  
The presence of the H$_2$ molecule is found to induce 
a small local non-superfluid helium 
density in the second solvation shell 
which makes a non-negligible contribution to the
moment of inertia of the complex in helium.  
The resulting moments of inertia for the OCS-H$_2$ 
complex embedded in a cluster of 63 helium atoms
are found to be in good agreement 
with experimentally measured values in large helium droplets. 
Implications for analysis of 
rotational constants of larger complexes of OCS 
with multiple H$_2$ molecules in helium 
are discussed.  
\end{abstract}

\pacs{36.40.-c, 36.40.Mr, 67.40.-w, 67.40.Yv}

\maketitle

\section{Introduction}
\label{sec:intro}

Helium droplets are known to provide an attractively convenient environment 
for synthesis of weakly bound, metastable, or other unusual complexes
\cite{toennies01}.  The low temperature of these droplets ($\sim 0.15 - 0.4$ 
K), their liquid nature, and their weak interaction with impurity species 
render them ideal matrix hosts for weakly bound complexes.  Formation of 
such complexes is facilitated by the now standard pick-up technique
\cite{goyal85} in which foreign species are absorbed by the droplet in 
a pick-up chamber.  Sequential use of such pick-up allows assembly of a variety 
of complexes and aggregates of atoms and molecules.  In recent years 
this synthetic potential of helium clusters 
has been demonstrated with synthesis of water clusters,\cite{nauta00} 
of metal clusters,\cite{bartelt96,federmann99}, of metal-adsorbate clusters,\cite{nauta01} 
and of chains of HCN, \cite{nauta99} 
all in the ultra-cold environment of a helium droplet. 
In several of these instances high 
resolution spectroscopy of the embedded clusters has shown that metastable 
isomers of conformations not seen for the corresponding clusters in the gas 
phase are formed in helium.  This indicates that although weak, the effect of 
helium solvation is non-negligible in determining the structure of some 
complexes, in particular those formed from polar constituents.

Complexes of molecules with molecular hydrogen constitute a special type of 
complex whose formation and behavior in helium droplets is currently of great 
interest.  The possibility of finding a superfluid state of molecular hydrogen 
provides an additional motivation beyond that of merely forming the complexes 
and analyzing the effect of helium solvation on their structure.  
Molecular hydrogen superfluidity is most likely to be found for the $j=0$ state
of H$_2$, {\it i.e.}, para-H$_2$, the lightest bosonic molecule of the hydrogen 
isotopes.  We shall restrict ourselves to para-H$_2$ in this work, 
and for simplicity refer to it as H$_2$.  
Helium will in all cases be taken as the boson isotope, $^4$He.
A number of significant experimental results
have been obtained for OCS(H$_2$)$_M$ complexes
in recent years.  Ref.~\onlinecite{grebenev00a} showed that for 
$M \geq 11$ there exist 
spectral anomalies which are consistent with the existence of a superfluid state
of the H$_2$ molecules. Existence of this 
superfluid state of molecular hydrogen in a nanoscale system has recently 
been confirmed by path integral calculations and shown to have an onset 
at a temperature of $\sim 0.3$ K for hydrogen clusters around OCS.\cite{kwon02}
Additional spectroscopic studies have been made for $M=1-8$ 
in helium droplets 
and analyzed in terms of rigidly 
coupled models of the OCS(H$_2$)$_M$ complexes.
\cite{grebenev01a,grebenev01} 
For the smallest $M=1$ complex 
of OCS-H$_2$,  the corresponding gas phase spectroscopy 
has recently been measured
and analyzed in Ref.~\onlinecite{mckellar02}.  
This smallest complex provides a reference point 
for theoretical analysis of this and all larger clusters 
with $M > 1$ in helium.

In this work we undertake a detailed theoretical analysis of 
the $M=1$ complex OCS-H$_2$ in droplets of $^4$He, using path 
integral methodology and the local two-fluid theory.\cite{kwon99}  
We make path integral Monte Carlo calculations of the complex in 
clusters of $N=63$ helium atoms.  This size is large enough to
provide a complete first solvation shell around the OCS molecule 
($\sim N=20$)\cite{paesani01c} that has a robust and compact structure 
which is independent of further increases in $N$.  
Both hydrogen molecule and helium atoms are described
quantum mechanically, with full permutation exchange 
symmetry of the helium component incorporated.  We compute the local 
non-superfluid helium solvation density induced by the OCS-He and 
H$_2$-He interactions.  Comparison of this non-superfluid density 
around OCS in pure $^4$He$_{64}$ and around OCS-H$_2$ in $^4$He$_{63}$ shows 
that the H$_2$ molecule induces a small non-superfluid density around itself
in the second solvation shell of helium as well as in the first shell.  

We evaluate the effective moment 
of inertia of the OCS molecule assuming that the H$_2$ molecule 
is rigidly coupled to the molecular rotation and employing the microscopic two-fluid theory of Ref.~\onlinecite{kwon99} to evaluate the contribution from the helium solvation density. 
The assumption of rigid coupling of H$_2$ to the molecular rotation 
is tested by an evaluation of the rotational constants 
of free (gas phase) OCS-H$_2$ and by comparison of these 
with the experimental values measured in Ref.~\onlinecite{mckellar02}. 
This comparison, made with interaction potentials\cite{higgins03} 
that have been previously calibrated for OCS-H$_2$ by exact bound state
calculations,\cite{zillich03a} shows that the 
free OCS-H$_2$ complex can be very accurately described with a rigid 
coupling model.   
This free complex is planar, with mass distribution corresponding to an 
asymmetric top. Excellent agreement of the calculated moments 
of inertia with experimental
values is found.  The free complex calculations are 
followed by a local two-fluid 
analysis of the response of the solvating helium density to
rotation of this OCS-H$_2$ complex, 
to obtain estimates for the helium contribution to all 
three moments of inertia in helium droplets. 
The average moment of inertia resulting from 
this analysis incorporating both first and second shell 
complex-induced non-superfluid helium densities   
is in excellent agreement with the corresponding 
experimentally measured value in helium 
droplets, with an accuracy of $\sim$ 4\%.   
The calculated inertial defect indicates 
a non-planar mass distribution of the total
helium-solvated complex, in agreement with the experimental measurements.
Comparison of individual moments of inertia 
with their corresponding experimental values shows somewhat larger deviations 
of 10 - 20\%.
The smaller accuracy of the individual moments of inertia 
can be ascribed to limitations 
of the local non-superfluid estimators used here 
in distinguishing accurately between rotations around different axes.

Such quantitative agreement of the moments of inertia and of the corresponding 
rotational constants with experimental measurements implies that the underlying
model of an OCS molecule rotating with a rigidly coupled H$_2$ molecule and 
a rigidly coupled local non-superfluid density of helium 
provides an accurate description for the low rotational states of 
the OCS-H$_2$ complex that are accessed by spectroscopic measurements. 
Given that high quality interaction potentials calibrated 
by experimental measurements of the free OCS-He and OCS-H$_2$ complexes 
are employed here, we can take the agreement between these results 
for the hydrogen complex of OCS embedded in helium with experiment 
as providing further evidence for the accuracy of the local two-fluid model 
of quantum solvation and superfluid response to rotational motion
around heavy molecules.\cite{kwon99,kwon00}

The structure of the rest of the paper is as follows. 
In Section~\ref{sec:method} we summarize the path integral methodology used 
for these calculations with both helium and hydrogen treated quantum 
mechanically.  In Section~\ref{sec:calibration} we describe the interaction 
potentials and present the various calibration calculations for the gas phase 
OCS-H$_2$ complex that provide a validation 
of our assumption that at low energies the dynamics of the H$_2$ molecule
are rigidly coupled to the molecular rotation. 
Section~\ref{sec:results} then presents the results 
of the full path integral calculations and 
microscopic two-fluid analysis for the OCS-H$_2$ complex 
when embedded in helium clusters.
We conclude in Section~\ref{sec:conclude} 
with a summary and brief discussion of implications 
for larger complexes of OCS(H$_2$)$_M$ in helium.

\section{Methodology}
\label{sec:method}

In this work 
the OCS molecule is treated as fixed and non-rotating.  
These restrictions are convenient because of the greater computational 
cost associated with the quantum 
treatment of the additional H$_2$ molecule. 
The consequences of these restrictions are not severe 
for the two-fluid analysis, as the following arguments indicate.
Previous studies have shown that for heavy molecules the helium solvation 
density around an embedded molecule is not significantly affected 
by the molecule translation motion,\cite{mcmahon96,huang02a}  
but that there is a significant decrease in the angular modulation 
of this solvation 
density when the molecular rotational degrees of freedom are incorporated.
\cite{paesani01c,patel02}  Very recent path integral studies that incorporate 
all molecular translational and rotational degrees of freedom have shown that 
the integrated local non-superfluid density is insensitive to this modulation
and that the moment of inertia contributions from the local non-superfluid 
may be accurately estimated from calculations made 
for a non-rotating and non-translating OCS.\cite{zillich03b}
Therefore we shall employ a non-rotating OCS molecule here and postpone 
investigation of the detailed effects of OCS rotational and translational 
motions to future work.

\begin{figure}
\resizebox{\columnwidth}{!}{\includegraphics{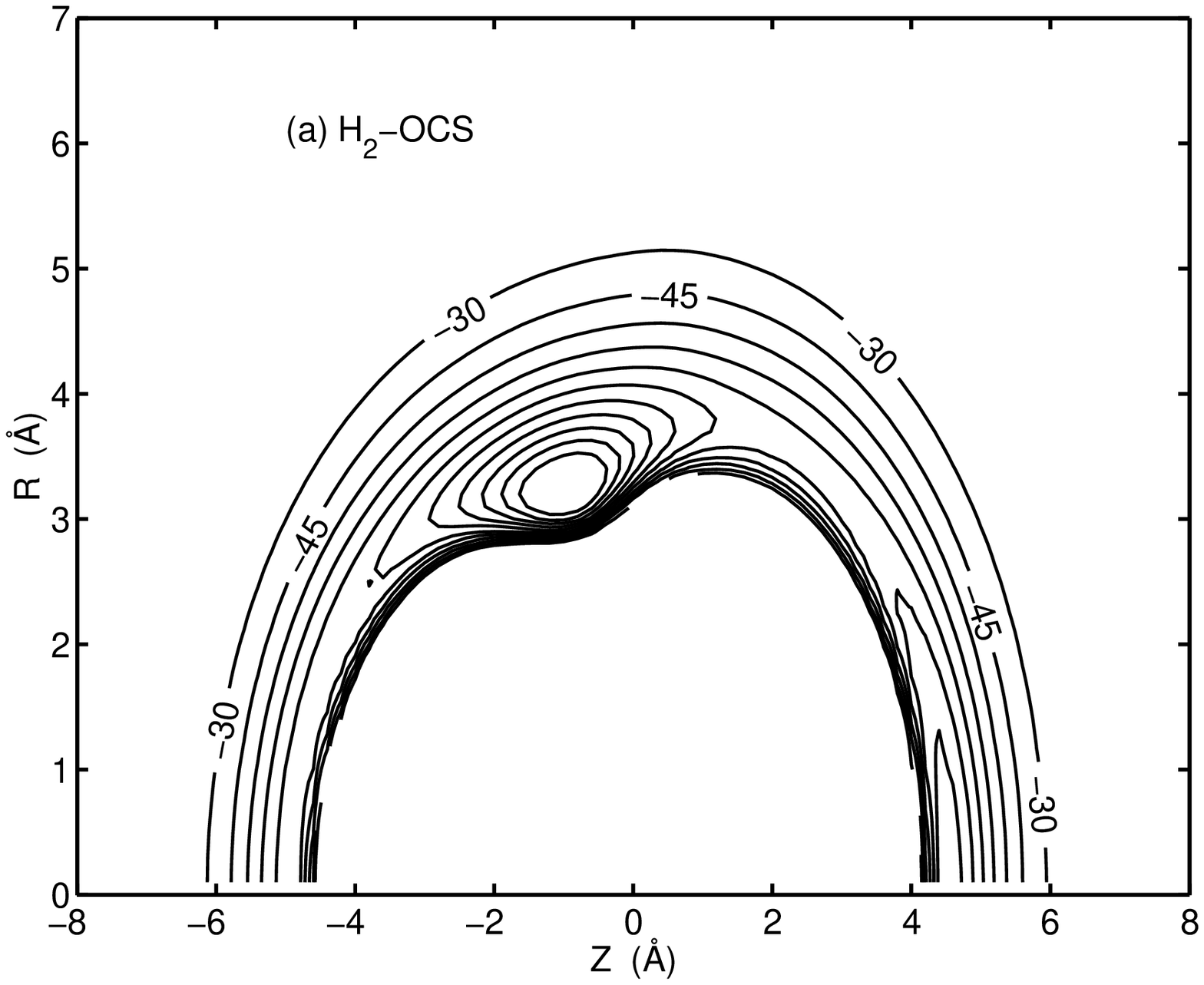}}
\resizebox{\columnwidth}{!}{\includegraphics{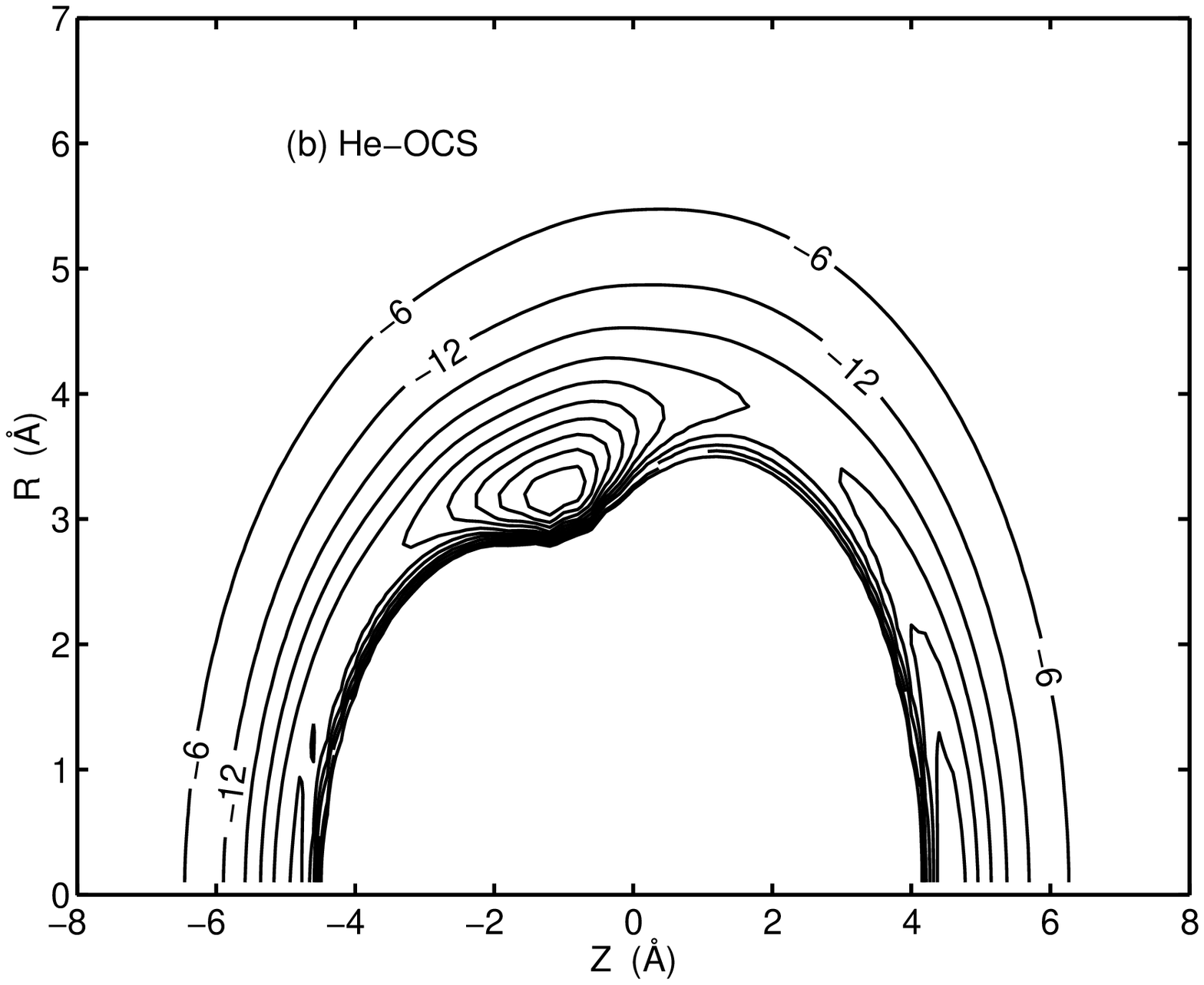}}
\caption{ Contour plots of the H$_2$-OCS and He-OCS potentials, 
shown as a function of the cylindrical coordinates ($Z$,$R$), 
where $Z$ is the coordinate along the axis passing through the OCS molecule 
and $R$ is the radial distance from the OCS molecular axis.
The origin is set at the center of mass of the OCS and the molecule
is oriented as O-C-S from $-Z$ to $+Z$.
Contours are shown in increments of 15 K for the H$_2$-OCS 
and 6 K for the He-OCS potential, respectively.
}
\label{fig:potentials}
\end{figure}

For the interaction potentials between the three different particles 
involved (para-H$_2$, He, and OCS),
we use a sum of pair potentials consisting of
H$_2$-He,\cite{bergh88} He-He,\cite{aziz87} OCS-H$_2$,\cite{higgins03} 
and OCS-He,\cite{higgins99} terms.
The para-hydrogen molecules are treated as spherical particles like
the helium atoms.  This is a well justified approximation 
at these low temperatures because of the
large rotational constant and nearly isotropic ground-state
electronic configuration of H$_2$. 
Consequently, we average the three-dimensional
ab initio OCS-H$_2$ potential 
which has recently been computed by Higgins {\it et al.}
with fourth-order M{\o}ller-Plesset (MP4) perturbation
theory,\cite{higgins03} over the orientation of the H$_2$ molecule.
This results in a two-dimensional representation of the OCS-H$_2$
interaction that has
a very similar topology to
the previously reported MP4 OCS-He potential.\cite{higgins99} 
Contour plots of the two potentials are shown in Figure~\ref{fig:potentials}. 
It is evident that the OCS-H$_2$ potential has its global minimum 
at approximately the same location as the OCS-He potential, 
but possesses a significantly deeper value.  
For OCS-H$_2$, the global minimum is $-208$ K, located at $r=3.35$~\AA~and
$\theta = 106^\circ$, where $r$ is the distance from the OCS center of mass
and $\theta$ the polar angle measured from the sulfur side
of the OCS molecular axis.  For OCS-He, the global minimum is 
$-65.3$ K and is located at 
$r=3.38$~\AA, $\theta = 108^\circ$. 
All calculations described here employ these two MP4 potentials for the interaction of OCS with H$_2$ and He.  For the H$_2$-He and He-He interactions
we use the empirical potential proposed by van den Bergh and Schouten,\cite{bergh88} and  
the semi-empirical potential of Aziz {\it et~ al.},\cite{aziz87} respectively.

We now have the following system Hamiltonian for the H$_2$-OCS complex
inside the $^4$He$_N$ cluster:
\begin{multline}
  H  = - \frac{\hbar^2}{2m_{\rm H_2}} \nabla^2_0
     - \frac{\hbar^2}{2m_{\rm He}} \sum_{i=1}^{N} \nabla^2_i 
    + \sum_{i<j} V_{\rm He-He}(r_{ij}) \\
    +  \sum_{i=1}^N V_{\rm H_2-He}(r_{0i})
    + V_{\rm H_2-OCS} (\vec{r}_0 )
    + \sum_{i=1}^N V_{\rm He-OCS} (\vec{r}_i ) .
\label{eq:hamiltonian}
\end{multline}
Here $m_{\rm H_2}$ ($m_{\rm He}$) is the mass of a hydrogen molecule (helium atom)
and $\vec{r}_0$ is the position vector of the hydrogen molecule in a (space fixed, for a non-rotating molecule) frame centered on the OCS molecule.

To study the OCS-H$_2$ complex and its helium solvation structure
we employ here the path integral Monte Carlo (PIMC) technique.  This approach allows us to make the quantitative analysis of
superfluidity in the helium environment that is required for the local 
two-fluid theory.   The main elements of the PIMC technique have been described
in Ref.~\onlinecite{ceperley95} and its adaptation to molecule-doped helium clusters 
detailed in Ref.~\onlinecite{huang02a}.  We give here only a brief summary, 
with methodological details that are specific to the OCS-H$_2$ and its helium surroundings.
The thermal density matrix is given by
\begin{equation}
\rho({\cal R},{\cal R}';\beta) \equiv \langle {\cal R} | e^{-\beta H} | {\cal R}' \rangle ,
\label{eq:dmatrix}
\end{equation}
with $\beta=1/k_B T$, 
${\cal R}$ a $3(N+1)$-dimensional vector, ${\cal R} \equiv ( \vec{r}_0,
\vec{r}_1, ...\vec{r}_{N} )$, 
and $H$ equal to the
Hamiltonian, Eq.~(\ref{eq:hamiltonian}), for an H$_2$ molecule
and $N$ helium atoms in the external
field provided by the stationary OCS molecule.
In order to incorporate the bosonic symmetry of the $^4$He atoms,
the density matrix is symmetrized by summing all permutations ${\cal P}$
amongst the $N$ helium atoms:
\begin{equation}
\rho_B({\cal R},{\cal R}';\beta) = \frac{1}{N!} \sum_{\cal P} \rho({\cal R},{\cal P}{\cal R}';\beta) .
\label{eq:bosedm}
\end{equation}
Since the density matrix of a interacting quantum system
is generally not known at a low temperature $T$, it is replaced in the Feynman path integral representation with a product of $L$ higher-temperature density
matrices, resulting in the discrete time path integral:
\begin{multline}
\rho({\cal R},{\cal P}{\cal R}';\beta) = \int \cdots \int d{\cal R}_1 d{\cal R}_2 \cdots d{\cal R}_{L-1} \\
\rho({\cal R}, {\cal R}_1 ;\tau) \rho({\cal R}_1,{\cal R}_2;\tau) \cdots \rho({\cal R}_{L-1},{\cal P}{\cal R}';\tau).
\label{eq:convol}
\end{multline}
Here $\tau=\beta / L$ constitutes the imaginary time step defining this discrete
representation of the path integral.
At sufficiently high temperatures $LT$, or equivalently, at sufficiently small time steps $\tau$, the density
matrix $\rho({\cal R},{\cal R}';\tau)$ can be approximated by a product of the free particle
propagator and an interaction term.
For the spherical He-He and H$_2$-He interactions, 
the pair-product form of the exact two-body density matrices can be used, and a matrix squaring methodology employed to achieve accurate grid representation of these.\cite{ceperley95}
The anisotropic He-OCS and H$_2$-OCS interactions are treated here
within the primitive approximation.\cite{huang02a}  
From our previous study of OCS-doped He$_N$ clusters,
we have found that a time step of $\tau^{-1}=80 K$ is 
required to get converged helium density profiles around the molecule
with this approximation for the high-temperature density matrix.\cite{kwon01}  
In order to make a simultaneous calculation of the sum over $N$-particle permutations in Eq.~(\ref{eq:bosedm}) together with
the multi-dimensional integration over the position coordinates in the discrete time path integral, Eq.~(\ref{eq:convol}), we employ
a stochastic process which samples the discrete paths
$\{ {\cal R},{\cal R}_1,{\cal R}_2, \ldots, {\cal R}_{L-1},{\cal P}{\cal R}'\}$ with probability density
proportional to
$\rho({\cal R},{\cal R}_1;\tau) \rho({\cal R}_1,{\cal R}_2;\tau) \cdots \rho({\cal R}_{L-1},{\cal P}{\cal R}';\tau)$.
This sampling is performed by the generalized Metropolis sampling algorithm of Pollock and Ceperley.\cite{pollock87}  A detailed description of this algorithm is provided in Refs.~\onlinecite{ceperley92},~\onlinecite{ceperley95} and ~\onlinecite{huang02a}. 
The thermal average of any observable $\hat{O}$ can then be determined by computing an arithmetic average of
$\langle {\cal R}' | \hat{O}| {\cal R} \rangle$ over the paths sampled.  

One of the advantages of using the path integral approach
is that it allows a quantitative estimate to be made of the superfluidity
of bosonic systems such as $^4$He and para-H$_2$ clusters as a function of temperature. 
Within the Feynman path integral analysis, 
the global superfluid fraction 
that is defined as the linear response of the bosonic system to classical rotation of its 
boundaries can be evaluated by an estimator written in terms of 
the projected area of the Feynman paths\cite{sindzingre89}:
\begin{equation}
f^s_{ij}= \frac{4m^2 \langle A_i A_j \rangle k_B T}{\hbar^2 I^{cl}_{ij}}.
\label{eq:area_est}
\end{equation}
Here $A_i$ is the area of a Feynman path projected onto a plane
perpendicular to the axis $\hat{x_i}$, and $I^{cl}$ is the classical
moment of inertia tensor.
This estimator yields a non-negligible value only when
exchange-coupled paths are comparable to the size of the system.
One problem that arises when applying this global superfluid estimator to analysis of doped helium clusters is that
it does not give information about the {\em local} perturbation
of superfluidity due to the presence of an impurity.  
However, it is precisely the local perturbation of the superfluid 
that determines the dynamic response of the helium to the rotation of 
an embedded molecule,\cite{kwon99,kwon00} so that a local analysis 
of superfluidity in the solvating helium density is required. 
Another possible issue with this estimator 
is that
it provides a measure of the superfluid response to classical rotation of a boundary 
as the rotation speed goes to zero, which does not map to the quantized rotation of
a molecule.  While this may not be very significant for analysis of the molecule-induced local
non-superfluid density, given that this appears to be independent of the effects of 
molecular rotation,\cite{zillich03b} it should be borne in mind when interpreting the 
meaning of the local superfluid density. 

In order to estimate the local superfluid
fraction around an impurity molecule, we 
compute here the density distribution
of helium atoms participating in permutation exchanges
which result in paths that are long compared to the system size.
In our previous studies of SF$_6$$^4$He$_N$ and OCS$^4$He$_N$ clusters, 
\cite{kwon99,kwon00,kwon01}
we found that a robust local superfluid distribution in the first
shell region 
could be computed for $N \ge 50$ 
by counting the exchange paths involving more than six
helium atoms.  The result is robust to the precise value of this cut-off, {\it i.e.}, similar results are obtained with five- or seven-atom exchange paths.   
As pointed out in Ref.~\onlinecite{kwon00}, this local estimator based purely on the exchange path length
does not reflect the tensorial nature of the superfluid response that is predicted from the linear response definition, Eq.~(\ref{eq:area_est}).   An alternative local estimator that is derived from decomposition of Eq.~(\ref{eq:area_est}) into contributions from small cells has recently been proposed in Ref.~\onlinecite{draeger03}.  This alternative estimator does possess the correct tensorial nature of the linear superfluid response, but 
unfortunately it is subject to extremely large fluctuations. To obtain meaningful values 
within small regions such as the first solvation shell of OCS would require an extremely 
high computational effort. (The application of this estimator in 
Ref.~\onlinecite{draeger03} 
employed cylindrical averaging over the length of a linear chain of three HCN molecules, 
which reduces the fluctuations.)  Thus, in order to make a microscopic analysis of the 
effect of the surrounding helium on the moment of inertia of the embedded H$_2$-OCS 
complex, we employ here the original local estimator of superfluid fraction employing 
long exchange path length.

\section{Structure and Rotational Constants of Free OCS-H$_2$}
\label{sec:calibration}

\begin{figure}
\resizebox{\columnwidth}{!}{\includegraphics{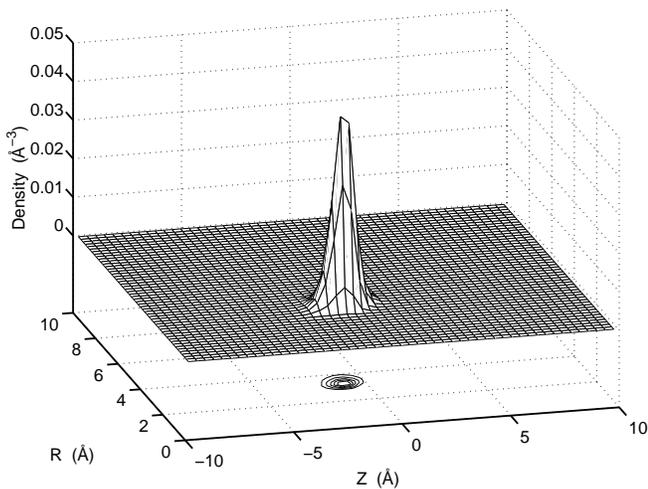}}
\caption{ Density of H$_2$ in the free OCS-H$_2$ complex, 
shown in the cylindrical coordinates $Z$ and $R$ with origin 
at the OCS center of mass (see Figure~\ref{fig:potentials}).
}
\label{fig:density_H2}
\end{figure}

We first performed a PIMC calculation on the free H$_2$-OCS complex,
{\it i.e.}, without any helium atom present.  Figure~\ref{fig:density_H2} shows
the density distribution of the single hydrogen molecule around OCS 
in the absence of helium.
The hydrogen molecule is clearly located at and around the global minimum 
of H$_2$-OCS potential. Its average position is $r=3.63\pm 0.33$~\AA,
$\theta = 106.5^\circ \pm 8.5^{\circ}$.
The uncertainties are evaluated as the fluctuations
$\Delta x = \sqrt{\langle x^2 \rangle - {\langle x \rangle}^2}$ 
which provide a measure of the widths of the single peak in the density 
distribution.  These PIMC values  of $r$ and $\theta$ extracted from 
the finite temperature density
distributions are very similar to the corresponding values obtained from 
ground state expectation values, namely  $r=3.704$ and $\theta=105.7$.
\cite{zillich03a} The average hydrogen position defines an average center 
of mass of the H$_2$-OCS complex.  
Transforming to the complex body-fixed frame, yields the average coordinates 
$r'=3.51\pm 0.33$~\AA~and $\theta' = 106.5^{\circ} \pm 8.5^{\circ}$, 
where $r'$ is the distance from the complex center of mass and $\theta'$ 
the polar angle from the complex body-fixed $z$-axis which is defined 
to be parallel to the OCS molecular axis.  
We see that there is only a very slight displacement 
of the center of mass away from the OCS center of mass 
($\sim 0.117$ \AA) as a result of complexing with the light hydrogen molecule. 

We compute the moment of inertia tensor of this complex 
assuming that the hydrogen 
is rigidly attached to the OCS molecular rotation.  
Thus, at each imaginary time configuration we evaluate 
the instantaneous moment of inertia tensor in the (instantaneous) 
body-fixed frame of the complex 
which is centered on the center of mass of the complex
with $z$-axis parallel to the molecular axis,
$x$-axis perpendicular to this and in the OCS-H$_2$ plane, and
$y$-axis mutually perpendicular to both $x$- and $z$-axes.
Averaging over the path integral yields 
the tensor components
\begin{equation}
I_{ij} = [I_0]{ij} +  m_{H_2} \langle ( r^2 \delta_{ij} - x_i x_j ) \rangle ,
\label{eq:moi_free}
\end{equation}
where $I_0$ is the gas-phase moment of inertia tensor of free OCS,
and the coordinates $x_i$ are the components of the hydrogen position vector 
$\vec{r}$ in the body-fixed frame.  
Note that the $y$-axis is a principal axis of the moment of inertia tensor
and corresponds to axis $\hat{c}$ below.
We then diagonalize this tensor to obtain the three principal 
values for the moments of
inertia, namely $I_a$, $I_b$, and $I_c$, and their corresponding principal axes $\hat{a}$ $\hat{b}$, and $\hat{c}$.  
We obtain the rotational constants of the complex from
the inverse of the principal moments of inertia, {\it e.~g.}, $A=\hbar^2/2I_a$.

\begin{figure}
\resizebox{\columnwidth}{!}{\includegraphics{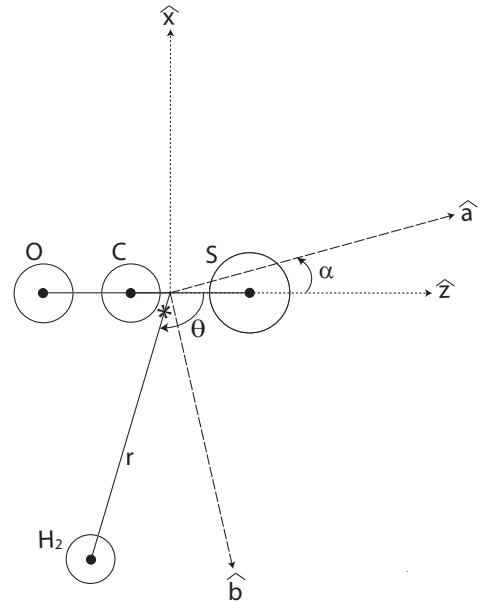}}
\caption{Schematic of the geometry and principal axes of 
the free OCS-H$_2$ complex. $\hat{x}$ and $\hat{z}$ denote 
the coordinate system in the body fixed frame centered on 
the OCS molecule center of mass.  
The principal axes ($\hat{a}, \hat{b}$) are shown here in this frame.
The third coordinate $\hat{y}$ and principal axis $\hat{c}$ are perpendicular
to the plane of the paper, with $\hat{y}$ directed out of the page 
and $\hat{c}$ into the
page.  The asterisk denotes the
complex center of mass, located at 0.117 \AA~away from the molecule center of mass
on the OCS-H$_2$ axis. Distances are shown to scale.
}

\label{fig:geometry}
\end{figure}

The resulting moments of inertia, $I_a$, $I_b$, and $I_c$, and rotational constants 
$A$, $B$, and $C$ are listed in Table~\ref{table1} (PIMC(1)).  The corresponding principal axes are: 
$\hat{a}$ in the $zx$-plane rotated an angle $\alpha$ away from the body-fixed $z$-axis, 
$\hat{b}$ perpendicular to this and within the same plane, and 
$\hat{c}$ perpendicular to the $zx$-plane. 
The angle $\alpha$ is also listed in Table~\ref{table1}.
Figure~\ref{fig:geometry} shows the relation between the body fixed 
and principal axes of the complex.
To assess the effect of the displacement of
the center of mass away from the OCS molecule, 
we have also calculated the moment of
inertia tensor and corresponding values of $A$, $B$, $C$, and $\alpha$ 
in a body-fixed 
frame that is centered at the OCS center of mass.  
These values are shown in Table~\ref{table1} 
as PIMC(2). The displacement of the center of mass away from the OCS axis 
is seen to have only
small effects on the rotational constants and the angle $\alpha$. 
The PIMC results are compared in Table~\ref{table1} with i) the values obtained
from exact
calculations employing close-coupling and descendant-weighted diffusion
Monte Carlo calculations with the same H$_2$-OCS potential\cite{zillich03a}, and
 ii) recent
experimental measurements.

\begin{table}
\caption{Moments of inertia (in amu \AA$^2$) and rotational constants (in cm$^{-1}$)
of the free OCS-H$_2$ complex. 
$\alpha$ is the angle between the OCS molecular axis and the $\hat{a}$ principal axis.
PIMC(1) is obtained with the body-fixed frame centered at the true
center of mass of the complex.  PIMC(2) is obtained with the 
approximation that the body-fixed frame is centered at the OCS center of mass.  The
experimental value of $\alpha$ was estimated from the measured intensity ratios in
Ref.~\onlinecite{mckellar02}, $\alpha=arctan(\mu_b/\mu_a)$.
The exact results in column 3 derive from rotational energy levels obtained from the 
BOUND program,\cite{bound} 
(rotation constants and moments of inertia) and from a
ground state average of the moment of inertia evaluated with exact densities from
diffusion Monte Carlo calculations ($\alpha$).\cite{zillich03a}}
\label{table1}
\begin{ruledtabular}
\begin{tabular}{c|cccc}
       & PIMC(1)    & PIMC(2)  & EXACT\cite{bound,zillich03a}        & Expt.\cite{mckellar02} \\ \hline
 $I_a$& 21.93(2)    & 22.84(2) & 22.29                         & 22.14        \\
 $I_b$& 85.87(2)    & 86.39(2) & 84.50                         & 84.25       \\
 $I_c$& 108.26(2)   & 109.24(2)& 109.28                        & 109.78       \\
 $\alpha$ & 5.63$^{\circ}$ & 6.29$^{\circ}$& 5.83$^{\circ}$ & 7.97$^{\circ}$ \\ 
      &             &          &                               &          \\
  $A$  & 0.7679(7)   & 0.742(3) & 0.7554 			& 0.7607 \\
  $B$  & 0.1961(1)   & 0.1941(5)& 0.1993 			& 0.1999 \\
  $C$  & 0.1556(1)   & 0.1537(5)& 0.1541 			& 0.1534 
\end{tabular}
\end{ruledtabular}
\end{table}

We discuss first the comparison with the exact calculations. 
This comparison allows us to 
assess the accuracy of the rigid coupling assumption for H$_2$.  The PIMC rotational 
constants are seen to be in excellent agreement ($\sim 1\%$)
with the values derived from converged energy levels obtained with the close-coupling 
BOUND program.\cite{bound}  Consequently we can conclude that the assumption that H$_2$ is 
rigidly coupled to the OCS molecular rotation is indeed very accurate for OCS-H$_2$.  
This is not the case for the analogous OCS-He complex, as is investigated with other 
methods and discussed in more detail in Ref.~\onlinecite{zillich03a}.  

The experimental measurements in the right hand column of Table~\ref{table1} derive from 
a recent experimental characterization of the free OCS-H$_2$ complex by infra-red spectroscopy.\cite{mckellar02} 
From spectral fits made to observed transitions between low-lying energy levels of the 
complex, 
it was determined that OCS-H$_2$ is an asymmetric rotor at low energies.
Table~\ref{table1} shows that the experimentally measured rotational constants (column 3) 
are in excellent agreement (within $\sim 0.5$\%) with the values obtained from the exact 
BOUND calculations (column 2). This confirms that the OCS-H$_2$ interaction potential 
used here\cite{higgins03} is very accurate, at least in the region around its global 
minima where the single hydrogen is predominantly located. 

We have now confirmed both
the accuracy of both the OCS-H$_2$ potential and the rigid 
coupling approximation for the OCS-H$_2$ complex. Given these two calibrations, we can proceed to directly compare the PIMC rotational constants 
(column 1, PIMC(1)) with the experimental values (column 3).  The agreement here is also 
excellent ($\sim 0.5 -3 \%$). 
The rotational constants $A$ and $C$, the angle $\alpha$, and the free OCS 
rotational constant ($b=0.20285$ cm$^{-1}$)~\cite{hunt85} can be used 
with the ball-and-stick model of Hayman et al.~\cite{hayman89} to
obtain an independent estimate of the structural parameters $r' \equiv r_{bs}$ 
and $\theta' \equiv \theta_{bs}$.
We note that the 
experimentally derived estimates for the ball-and-stick parameters,
$r_{bs}=3.719$ \AA, $\theta_{bs}=110.8^{\circ}$,\cite{mckellar02} 
lie within the width of the hydrogen density peak 
in Figure~\ref{fig:density_H2}.  

The consistently close 
agreement of the PIMC results for all spectroscopic and structural quantities with the corresponding experimental quantities
therefore allow us to conclude that a rigid coupling analysis made with path 
integral densities for H$_2$ provides a highly accurate description 
of the free OCS-H$_2$ complex. 
We note also that the quantum averages are consistent with a simple 
ball-and-stick model, although the large size of the quantum fluctuations 
render this model of 
limited applicability beyond confirming the average structures.

\section{Structure and Rotational Constants of OCS-H$_2$ embedded in Helium}
\label{sec:results}

\begin{figure}
\resizebox{\columnwidth}{!}{\includegraphics{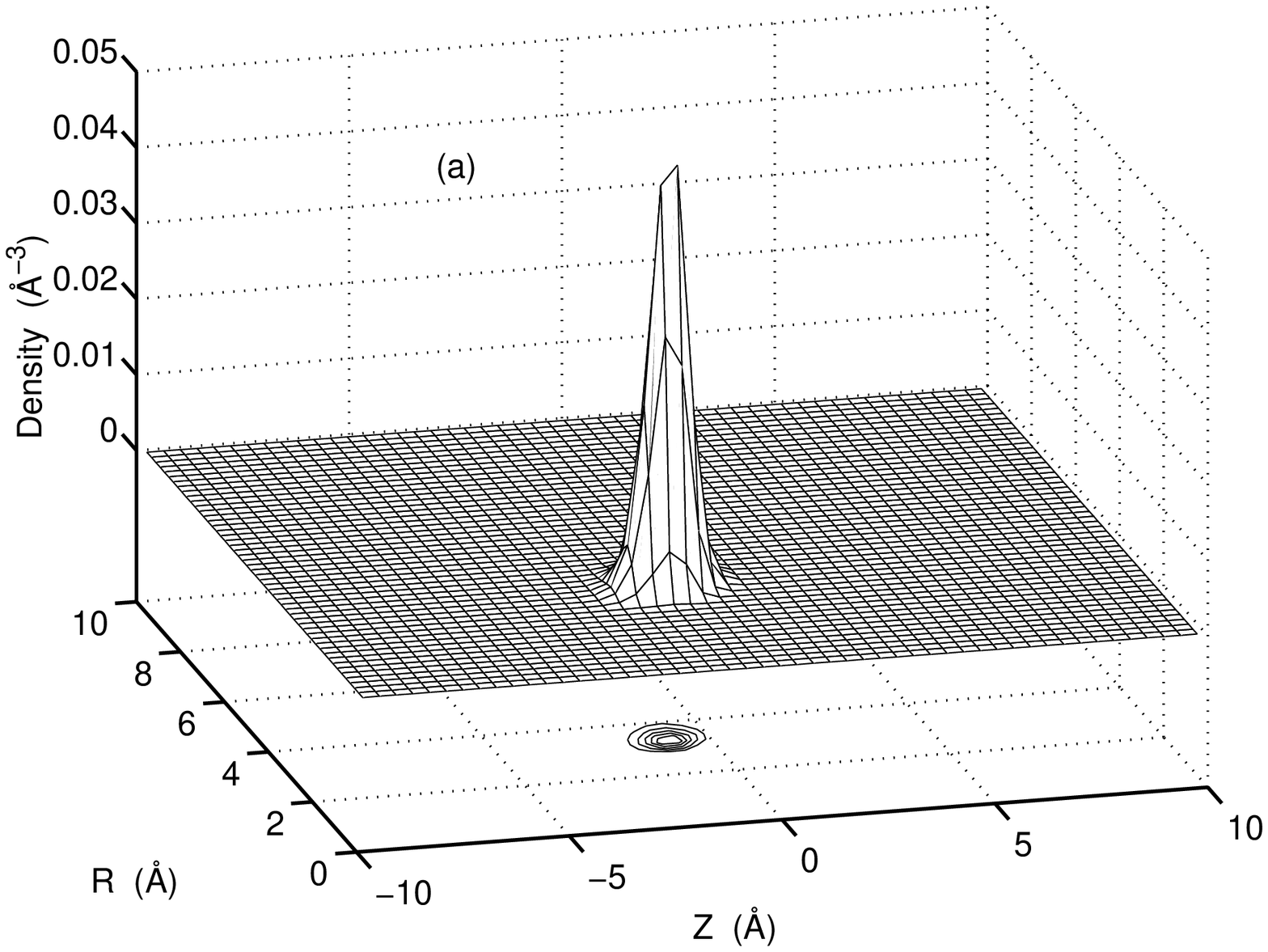}}
\resizebox{\columnwidth}{!}{\includegraphics{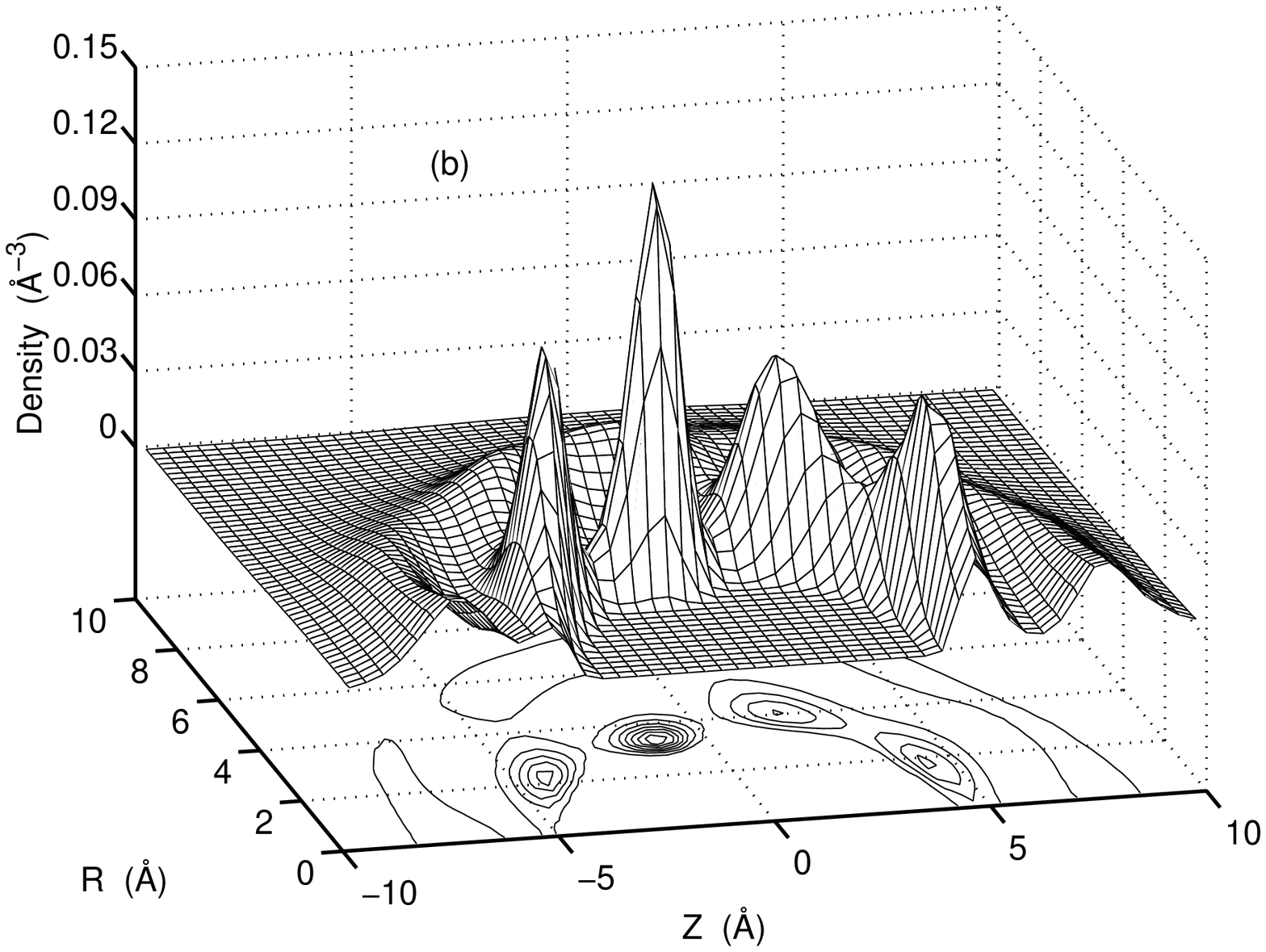}}
\caption{ Density of H$_2$ and He 
in the (OCS-H$_2$)$^4$He$_{63}$ cluster,
shown in the cylindrical coordinates $Z$ and $R$ with origin
at the OCS center of mass (see Figure~\ref{fig:potentials}).
}
\label{fig:density_HeH2}
\end{figure}

Figure~\ref{fig:density_HeH2}a shows the density distribution of 
the single H$_2$ when the 
OCS-H$_2$ complex is now embedded in a cluster of $N=63$ helium atoms.  
The corresponding helium density is shown in Figure~\ref{fig:density_HeH2}b.  
Note that the origin in all figures is set to be at the OCS center of mass.  
The hydrogen distribution is seen to
change very little as a result of the surrounding helium atoms 
(compare Figure~\ref{fig:density_HeH2}a 
with Figure~\ref{fig:density_H2}), except that the single density peak
is somewhat sharper in the presence of helium. This reflects that
the presence of the neighboring helium atoms makes the binding of the hydrogen
to the OCS more compact.
The average location of the H$_2$ molecule is now $r=3.59\pm 0.31$~\AA, 
$\theta = 105.8^\circ \pm 7.5^{\circ}$, very similar to its location 
in the free complex.  This location compares well with the values 
$r=3.9\pm 0.2$~\AA, $\theta = 111^{\circ}\pm 6^{\circ}$ extracted 
from the spectroscopic measurements in Ref.~\onlinecite{grebenev01a}. 
The close similarity of the structure of the OCS-H$_2$ complex with and 
without solvating helium suggests that the rigid coupling of the hydrogen 
to OCS remains a valid approximation when the complex is embedded in helium. 

Looking closely now at the helium distribution, 
it is evident that the first shell structure of this
looks very similar to the corresponding first shell structure for the 
OCS$^4$He$_{64}$ cluster (see Figure 3
of Ref. \onlinecite{kwon01}), with one significant difference.  
This is that the most intense density peak
at the global minima consists now of only five helium atoms, 
instead of the six helium atoms seen for OCS$^4$He$_{64}$.  
This difference is due to the fact that one
helium atom is now replaced by a more strongly-bound H$_2$ molecule. 
Related calculations with larger numbers of hydrogen molecules have shown 
that the six helium atoms located at the global minimum in OCS$^4$He$_N$ are 
completely replaced by five hydrogen molecules.\cite{patel01a} 
(Similar conclusions have been recently reached from an intensity analysis 
of spectral lines in Ref.~\onlinecite{grebenev02}.)

In order to estimate the rotational constants of the OCS-H$_2$ complex
inside the larger helium droplets, we use the microscopic two-fluid
theory of the response of the solvating helium density to the molecular
rotation.\cite{kwon99,kwon00} First, we investigate the local perturbation
of helium superfluidity around the OCS-H$_2$ complex, using the local superfluid
estimator based on the exchange length of the Feynman paths. 
As in our previous study for the SF$_6$- and OCS-doped helium droplets,\cite{kwon99}
we choose an exchange length of six or greater as the criterion for 
paths contributing to local superfluidity at any given position.  
The results are relatively insensitive to this cutoff length, 
since the paths are generally split between very small paths (2-5 exchanges) 
and very long paths ($\sim N$ exchanges). 

The moment of inertia tensor of the complex inside the helium droplet 
is then evaluated by summing the rigidly-coupled single hydrogen
and the solvating helium contributions to the gas-phase moment of inertia
tensor of OCS:
\begin{equation}
  I_{ij} = [I_0]_{ij} + [I_{H_2}]_{ij} + [I_{He}]_{ij} , 
\label{eq:moi_complex}
\end{equation}
where the hydrogen contribution $I_{H_2}$ is given by the second term 
in Eq. (\ref{eq:moi_free}). 
We employ here the body-fixed frame of the free OCS-H$_2$ complex 
defined with origin on the OCS center of mass 
(see Section \ref{sec:calibration}).  This is consistent with 
the reference frame employed in the analysis of the experimental 
measurements in Ref.~\onlinecite{grebenev01a}.

Within the local two-fluid approach the helium contribution can be estimated by
\begin{equation}
  [I_{He}]_{ij} = m_{He} \int \rho_{ns}(\vec{r})
                     (r^2 \delta_{ij} -x_i x_j) d\vec{r},
\label{eq:moi_contrib2}
\end{equation}
where $\rho_{ns}(\vec{r})$ is the local helium non-superfluid density.  
This is defined as
\begin{equation}
\rho_{ns}(\vec{r})= \rho(\vec{r}) - \rho_s (\vec{r}), 
\label{eq:nonsuperfluid}
\end{equation}
where $\rho(\vec{r})$ is the total number density of helium
and $\rho_s (\vec{r})$ the local superfluid density estimate obtained
by using the long exchange length criterion.
In our previous studies of the rotational constants
of impurity molecules such as OCS inside pure helium-4 clusters,
the molecule-induced non-superfluid fraction is localized essentially in the
first shell region.\cite{kwon99,kwon00} Any non-superfluid density in the second shell
was seen to result from cluster finite size effects and 
decreases as the cluster size
increases.\cite{kwon99}
However, in this case of the OCS-H$_2$ complex, the presence of the hydrogen molecule may 
induce
an additional non-superfluid fraction in the second shell region around the OCS.  We can determine the extent of this effect by comparing the local non-superfluid helium density in the vicinity of the H$_2$ molecule for 
(OCS-H$_2$)$^4$He$_{63}$ with the local non-superfluid density 
in the corresponding regions for OCS$^4$He$_{64}$.
\begin{figure}
\resizebox{\columnwidth}{!}{\includegraphics{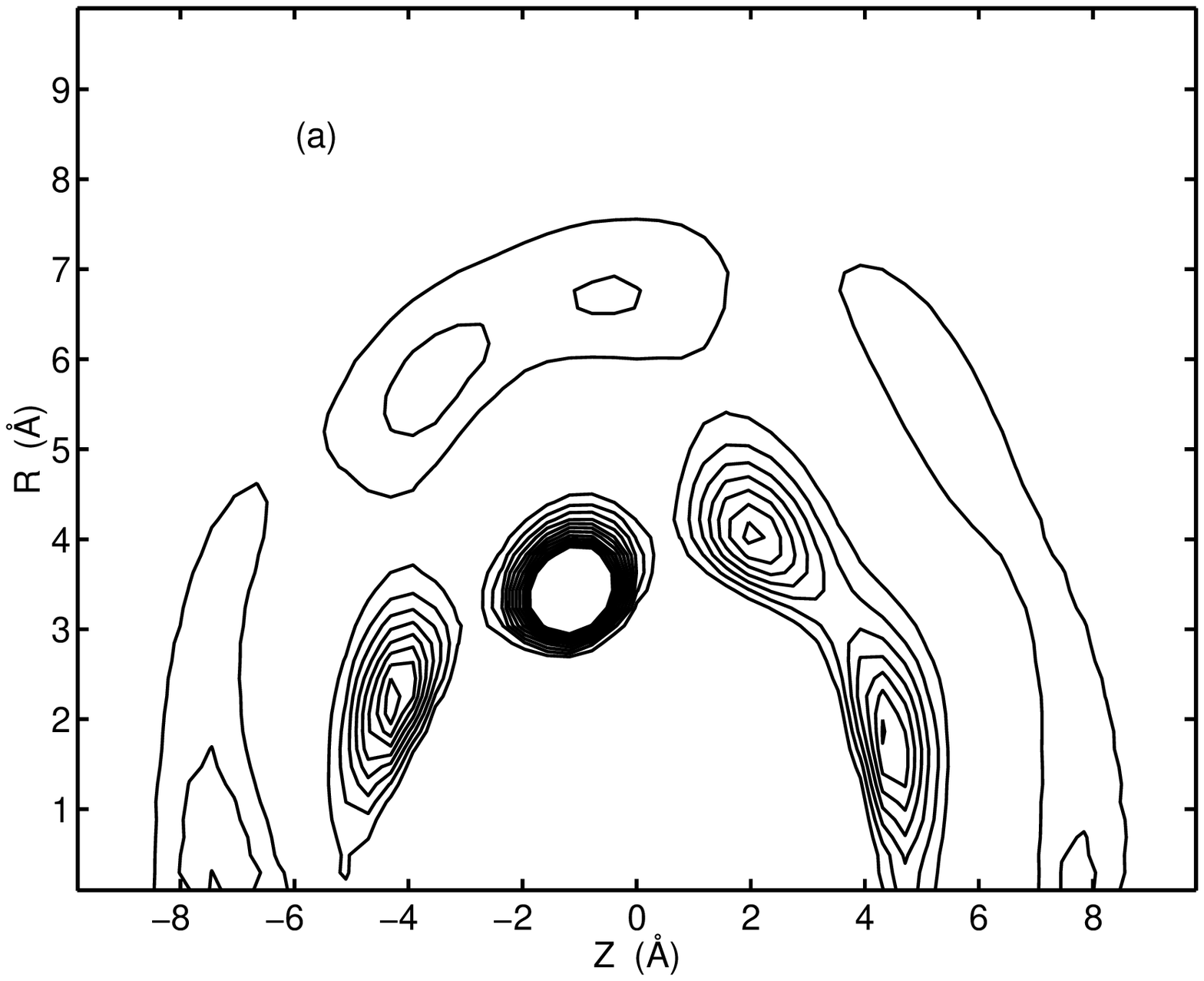}}
\resizebox{\columnwidth}{!}{\includegraphics{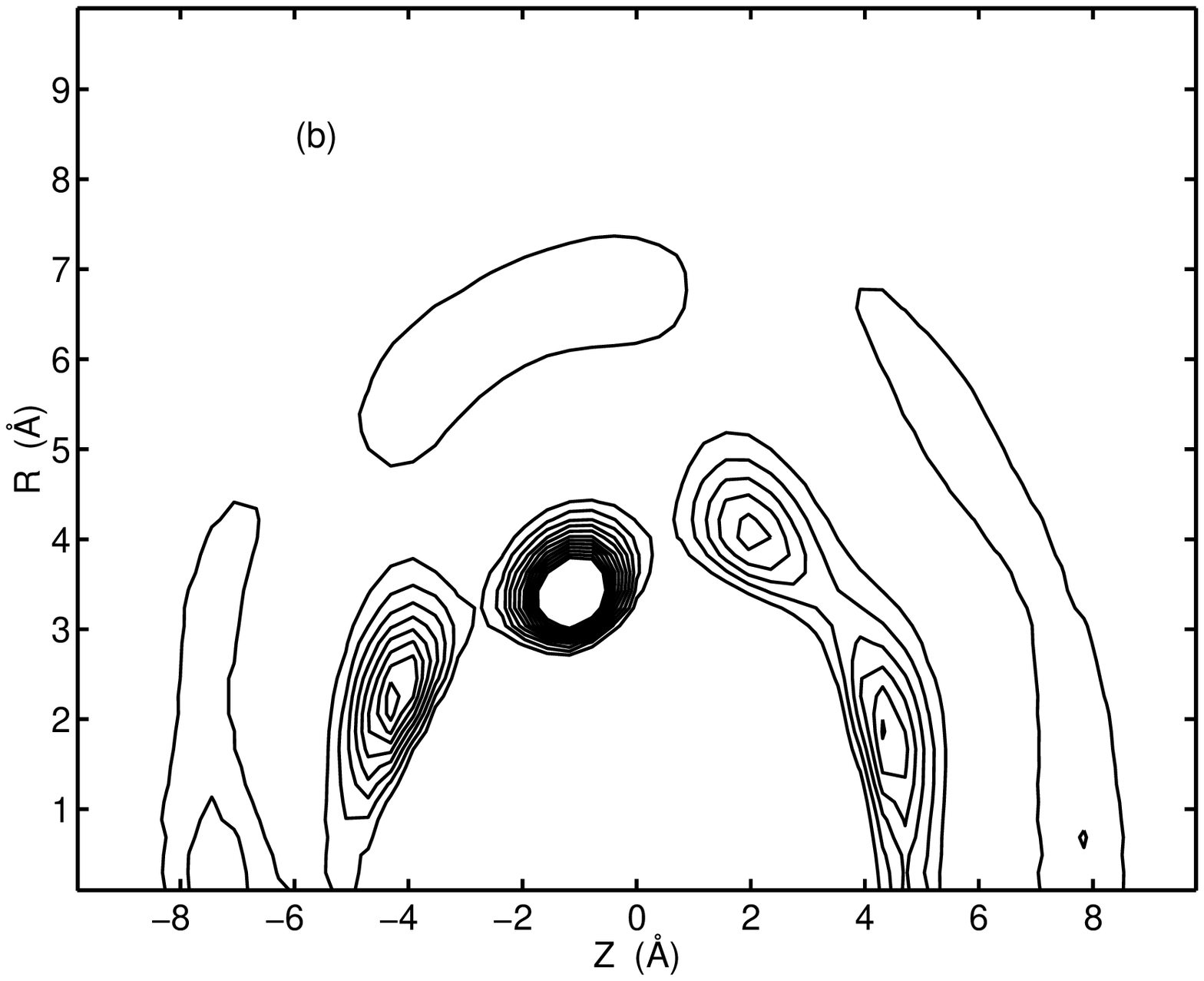}}
\caption{ Non-superfluid helium density around 
(a) the OCS-H$_2$ complex 
in (OCS-H$_2$)$^4$He$_{63}$, and around (b) OCS in OCS$^4$He$_{64}$. 
The densities are shown in mesh plots with contours below, 
in the cylindrical coordinates $Z, R$ defined in Figure~\ref{fig:potentials}.}
\label{fig:2nd_nonsf}
\end{figure}
Figures \ref{fig:2nd_nonsf} (a) and (b) show the non-superfluid helium density
distributions of $N=63$ helium atoms around
the OCS-H$_2$ complex and 
of $N=64$ helium atoms around the OCS molecule, respectively.
There is a clear increase in the non-superfluid density
in the second helium solvation shell around the OCS-H$_2$ complex in the vicinity of the H$_2$ molecule.  This small region of second shell 
non-superfluid density is evidently due to 
the presence of the distinguishable hydrogen molecule nearby.

In our two-fluid estimate of the effective moment of inertia,
both this additional non-superfluid fraction in the second shell
and the non-superfluid fraction of the helium 
in the first solvation shell are assumed to be rigidly coupled
to the rotation of the complex.
For computational convenience, we further assume that
the additional non-superfluid fraction in the second shell
lies in the instantaneous plane including the linear OCS molecule and the
single hydrogen molecule.  
This allows the second shell contribution to be estimated from 
integration over the density difference between Figures 
\ref{fig:2nd_nonsf} (a) and (b). Note that this provides 
an azimuthally averaged value 
for the second shell non-superfluid density. More realistically, 
we expect this density 
to be localized in three dimensions around the instantaneous 
H$_2$ azimuthal position.
Such a three-dimensional distribution will not necessarily 
be accurately described by an azimuthal average.
However, sampling the relative azimuthal positions of the H$_2$ and
local non-superfluid helium density is computationally extremely expensive, 
so we have restricted ourselves to the azimuthally averaged approximation here,
bearing in mind that this may lead to some inaccuracy
in the second shell contribution to the moment
of inertia tensor because of the inaccuracy of the underlying mass distribution.
The effective moment of inertia tensor of Eq. (\ref{eq:moi_complex}) 
is then diagonalized to yield 
the principal axes of the hydrogen complex in helium, $\hat{a}, \hat{b}$, 
and $\hat{c}$, and the principal moments of inertia tensor 
$I_a$, $I_b$, and $I_c$. 
 
Possible contributions from the superfluid are limited 
by angular momentum and adiabatic following 
constraints, as discussed in detail in Ref.~\onlinecite{kwon00}, 
and are consequently
very small.  A recent angular-momentum-consistent calculation 
for bare OCS in helium has confirmed that 
the superfluid contribution is negligible in this case,\cite{huang03} 
and this is expected to be the same for the OCS-H$_2$ complex.

The resulting moments of inertia and principal axis angle $\alpha$ are listed in Table~\ref{table2}.
\begin{table}
\caption{PIMC moments of inertia (in amu \AA$^2$), 
and principal axis angle $\alpha$
of the OCS-H$_2$ complex 
inside a $^4$He$_{63}$ cluster, compared with the corresponding
experimental values obtained from spectroscopic measurements
in large helium droplets ($N \sim 10^3$).\cite{grebenev01a} }
\label{table2}
\begin{ruledtabular}
\begin{tabular}{c|cc}
      & PIMC &  Expt. \\ \hline
  $I_a$& 246(12)   & 199(1)         \\
  $I_b$& 328(14)   & 297(3)         \\
  $I_c$& 352(14)   & 399(6)         \\
  $(I_a + I_b + I_c)/3$ &  309(13)  &  297(4)    \\
       &     &         \\                             
  $\alpha$ & 7.5$^{\circ}$ & 39(2)$^{\circ}$ \\
\end{tabular}
\end{ruledtabular}
\end{table}
The average of the three PIMC principal moments of inertia
is in excellent agreement with the experimentally derived average (to 4\%), 
lying within the statistical error of the path integral estimate.  
However we note that the PIMC estimates for three principal values
of the moment of inertia differ from their
corresponding experimental values $I_a$, $I_b$, and $I_c$, by $+23\%$, $+10$\% and $-11$\%,
respectively.  
There are two possible contributions to these individual discrepancies.  
First, as discussed above  
the local superfluid estimator based on exchange path length does not reflect 
the tensorial nature of the superfluid response, 
and consequently it may not distinguish effectively
among the three principal axes ${\hat a}$, ${\hat b}$ and ${\hat c}$.  
Second, as noted above, 
we have approximated the contribution from the second helium
solvation shell as deriving from a non-superfluid density  
restricted to the OCS-H$_2$ plane. Small inaccuracies in the second shell 
non-superfluid mass distribution resulting from this approximation may 
give rise to errors in the estimated contribution of this shell 
to the moment of inertia tensor.
The fact that the individual principal
moment of inertia values are not achieved at the same accuracy as
the average value, indicates that the mass distribution in the second shell
is not predicted with sufficient accuracy from our simple estimate
based on the azimuthally averaged non-superfluid density.
The difference between the experimental value of the principal axis 
angle $\alpha$ and the PIMC estimate of this seen in Table~\ref{table2} can also be 
understood 
to result from this effect, since like the individual moments of inertia, 
$\alpha$ derives from diagonalization of the moment of inertia tensor.
Finally, any inaccuracy in the individual values of $I_a$, $I_b$, and $I_c$ will
propagate to the
estimate of the inertial defect $\Delta = I_c - I_a - I_b $.  We obtain a value
$\Delta = -222$ amu \AA$^2$. 
This is negative, in agreement with the experimental
finding in Ref.~\onlinecite{grebenev01a} 
and implying a non-planar effective complex
in helium,\cite{gordy} but is larger in magnitude than the 
experimental value $\Delta \sim -100$ amu \AA$^2$.  
This discrepancy is also consistent with 
the 10 - 23\%~inaccuracy in the individual moment of inertia values 
deriving from the effect of using an approximate mass distribution from the second helium solvation shell. 

\begin{table}
\caption{The moments of inertia (in amu \AA$^2$) contributions
from the rigidly-coupled hydrogen (OCS-H$_2$), 
the first-shell non-superfluid of helium (He I),
and the second-shell non-superfluid of helium (He II), respectively. 
}
\label{table3}
\begin{ruledtabular}
\begin{tabular}{c|ccc}
                   &  OCS-H$_2$  &     He I      &  He II \\ \hline 
  $I_a$            &   22.82     &     182       &   41 \\
  $I_b$            &   86.10     &     219       &   23 \\
  $I_c$            &  108.92     &     200       &   43 \\
 $(I_a+I_b+I_c)/3$ &   72.61     &     200       &   36 \\
\end{tabular}
\end{ruledtabular}
\end{table}

In order to assess the importance of the second-shell helium contribution
to the moment of inertia of the complex,
we show in Table \ref{table3} the various contributions
to the moments of inertia, namely, from the H$_2$ density and
from the non-superfluid helium density in the first and second solvation shells,
respectively.  
As can be seen, the second-shell helium contribution is quite significant,
amounting to $\sim$ 18\% of the first-shell helium contribution to the average moment
of inertia of the complex.
We find that if the contributions from the second solvation shell
are omitted (OCS-H$_2$ column plus He I column in Table \ref{table3}), the average moment
of inertia differs from the experimental quantity by a greater amount, namely $-8$\%, while the individual principal moments of inertia 
differ now from experiment by $+3$\%, $+2$\% and $-23$\%, respectively. 
The fact that the average moment of inertia is now significantly lower and in poorer agreement with the corresponding experimental value
confirms the need to take the second shell non-superfluid contribution
into account. 

\begin{table}
\caption{PIMC rotational constants (in cm$^{-1}$) of the OCS-H$_2$ complex
inside a $^4$He$_{63}$ cluster, compared with the corresponding
experimental values obtained from spectroscopic measurements
in large helium droplets ($N \sim 10^3$).\cite{grebenev01a} }
\label{table4}
\begin{ruledtabular}
\begin{tabular}{c|cc}
      & PIMC &  Expt. \\ \hline
  $A$  & 0.069(4)  & 0.0847(4) \\
  $B$  & 0.051(3)  & 0.0567(6) \\
  $C$  & 0.048(3)  & 0.04226(6) \\
  $(A+B+C)/3$ & 0.056(3) & 0.04948(6) \\
\end{tabular}
\end{ruledtabular}
\end{table}

Since we calculate directly the moment of inertia tensor 
of the OCS-H$_2$ complex, a comparison of the principal moments of inertia
with corresponding experimental values is the best way 
for us to calibrate our accuracy.  
Assessment of the resulting rotational constants is also possible,
but is complicated by the need to invert the principal values and 
the consequent increase in error. 
Table \ref{table4} shows the comparison of the PIMC rotational constants
and the corresponding experimental values.
Even though the average rotational constant is less accurate 
than the average principal moment of inertia,
the PIMC-based estimate for the average rotational constant is still 
in good agreement with the experimental value, to $\sim$ 15\%.  Given the approximate 
description of the second shell helium contribution, this would appear to be excellent 
agreement.  

\section{Discussion} \label{sec:conclude} 
We have presented microscopic quantum calculations of the rotational constants of 
OCS-H$_2$ both as a free complex and when embedded in helium droplets.  The analysis was 
based on path integral calculations, and employed an assumption of rigid coupling of the 
H$_2$ motion to the OCS rotation, 
as well as the microscopic local two-fluid theory to 
evaluate the helium contribution to the moment of inertia.  The 
OCS-H$_2$ complex is found to be an asymmetric rotor, and 
to possess a very similar structure when free and when embedded in helium. 
The complex is seen to induce a local non-superfluid helium density 
in the second solvation shell, in the immediate vicinity of the H$_2$ molecule,
as well as in the first shell. 
The overall effect of the solvating helium is to change 
the complex geometry as measured
spectroscopically, from planar to non-planar, 
reflecting an additional moment of inertia
contribution from the solvating helium.
The calculated moments of inertia derived from the two-fluid estimates 
and corresponding
rotational constants are in excellent agreement with experimentally 
measured values for the free OCS-H$_2$ complex, 
with individual values within 3\%.  
For the OCS-H$_2$ 
complex embedded in a helium cluster, the average moment of inertia is in  
excellent agreement with the corresponding experimental average, within $4\%$, 
when the contribution from the non-superfluid density induced by 
the H$_2$ molecule in the second solvation shell of helium is taken 
into account.  The individual values of the moments of inertia have 
larger deviations from experiment, $10 - 23$\%, 
but are still in good agreement.
These individual deviations, as also the deviation in angle 
$\alpha$ and in magnitude of inertial defect $\Delta$
may derive from inaccuracies in the components of 
the moment of inertia tensor resulting 
from treating the second shell helium contribution as azimuthally symmetric. 

The two-fluid analysis made here provides a physically consistent explanation 
of the anomalous effective masses assigned to the H$_2$ molecule in 
the extended shell model proposed in Ref.~\onlinecite{grebenev01a}.  
In that model, an extension of the earlier 'donut' model for OCS in helium 
droplets,\cite{grebenev00b} the increased moment of inertia was assigned to 
contributions from H$_2$ and from non-superfluid helium in the first 
solvation shell only.  The latter was estimated from a close packing model 
of our PIMC helium density distributions,\cite{kwon00} with effective helium 
masses of 0.55 amu assigned to approximate 
the theoretical non-superfluid density.  
The contribution from the H$_2$ molecule was described with an adjustable 
effective mass parameter, whose best fit value to experiment was 10 amu.
\cite{grebenev01a}  This large increase in effective mass 
of H$_2$ was attributed 
in Ref.~\onlinecite{grebenev01a} to the substitution of one indistinguishable 
helium atom by a distinguishable H$_2$ molecule, without any associated change 
in the surrounding helium density distribution.  
The analysis made here provides 
a simpler and more physical explanation for 
the anomalously large effective mass 
of H$_2$, namely, that this necessarily results when the H$_2$-induced second 
shell non-superfluid density is neglected 
and the moment of inertia increment is assumed 
to result only from the first solvation shell density components.  
The fitted value of the
H$_2$ effective mass is larger than its natural mass because this adjustable 
parameter must absorb the second shell contributions 
in the extended shell model 
of Ref.~\onlinecite{grebenev01a}.  It is therefore identified as 
an artifact of the model that would be removed 
if the model were extended to include 
the second shell helium contribution.
   
A rigid coupling of the hydrogen molecule to rotation 
of a heavy molecule such as OCS seen here is 
not unexpected.  Related studies have shown that 
this description becomes less accurate 
for higher total angular momentum states of the complex.\cite{zillich03a}  
It is useful 
to compare this behavior of the OCS-H$_2$ complex with that of its cousin, 
the OCS-He complex, which is considerably less rigid (see, {\it e.g.}, 
the comparative analysis in 
Ref.~\onlinecite{zillich03a}). 
Despite having similar topology of interaction potentials, 
the H$_2$ molecule has considerably stronger binding to OCS than does He.  
This results in 
a more sharply modulated structure, less delocalization, 
and more rigidity in coupling 
of its density to the OCS motion, reflecting the presence 
of low lying excitations of a rigid complex.
Nevertheless, it is important to realize 
that the hydrogen component is still a highly 
quantum element and this becomes very evident for larger $M$ (see below).

The excellent agreement with a rigid coupling model 
for the free complex suggests that similar rigid coupling, 
or more general quasi-adiabatic coupling,\cite{quack91,zillich03a} 
descriptions might 
provide good approximations for the analysis of OCS(H$_2$)$_M$, 
both free and embedded in 
helium droplets.  However, at larger sizes, 
in particular for $M > 11$, we have recently
shown that the hydrogen component possesses an anisotropic 
superfluid state.\cite{kwon02}
This state is by definition characterized 
by a lack of rigid response to rotation about 
the molecular axis. The first solvation shell is complete at $M=17$, and
at intermediate $M$ values one can identify precursors 
to the superfluid state that 
possess significant 
permutation exchanges within certain segments of the local solvation 
structure.\cite{kwon02}  
Thus for intermediate sizes one expects the values of rotational
constants to reflect a balance between rigidity due to the strong OCS-H$_2$ and 
H$_2$-H$_2$ interactions, and lack of rigidity due to 
onset of these permutation exchanges.  The intermediate-size hydrogen complexes
with OCS provide a rich opportunity to analyze 
the detailed evolution from full rigidity to 
anisotropic rigidity and superfluidity.

The high accuracy of this two-fluid analysis for the moments of inertia 
and rotational constants of the 
OCS-H$_2$ complex when embedded in helium provides confirmation 
of the accuracy of the 
underlying local two-fluid theory.  This analysis of moments of inertia 
and rotational constants for OCS-H$_2$ inside a helium droplet 
provides the first microscopic quantum calculation of rotational constants 
for a van der Waals complex 
inside helium.  The average values are found to be highly accurate, 
and the slightly larger deviations for the individual values understandable 
in terms of simplifications of the helium non-superfluid density made here 
to allow computational tractability.  The present analysis thereby 
extends the previous findings of highly accurate two-fluid 
calculation of rotational constants for the isolated molecules SF$_6$ and 
OCS in helium clusters\cite{kwon99,kwon00} to embedded molecular clusters. 
In the case of these heavy molecules and clusters for which high quality 
interaction potentials are available 
to enable meaningful microscopic calculations to be made, the 
two-fluid theory thus appears to provide an 
accurate microscopic description of response of 
a solvating superfluid helium droplet to 
the molecular rotation.

\section{Acknowledgments}
This work has been supported by the Korea Science
\& Engineering Foundation through its Basic Research Program
(grant R01-2002-000-00326-0 to YK) and by the Chemistry
Division of the National Science Foundation (grant CHE-0107541 to KBW). 
KBW thanks the Miller Institute for Basic Research in Science
for a Miller Research Professorship for 2002--2003.


\begin{thebibliography}{39}
\expandafter\ifx\csname natexlab\endcsname\relax\def\natexlab#1{#1}\fi
\expandafter\ifx\csname bibnamefont\endcsname\relax
  \def\bibnamefont#1{#1}\fi
\expandafter\ifx\csname bibfnamefont\endcsname\relax
  \def\bibfnamefont#1{#1}\fi
\expandafter\ifx\csname citenamefont\endcsname\relax
  \def\citenamefont#1{#1}\fi
\expandafter\ifx\csname url\endcsname\relax
  \def\url#1{\texttt{#1}}\fi
\expandafter\ifx\csname urlprefix\endcsname\relax\def\urlprefix{URL }\fi
\providecommand{\bibinfo}[2]{#2}
\providecommand{\eprint}[2][]{\url{#2}}

\bibitem[{\citenamefont{Toennies et~al.}(2001)\citenamefont{Toennies, Vilesov,
  and Whaley}}]{toennies01}
\bibinfo{author}{\bibfnamefont{J.~P.} \bibnamefont{Toennies}},
  \bibinfo{author}{\bibfnamefont{A.~F.} \bibnamefont{Vilesov}},
  \bibnamefont{and} \bibinfo{author}{\bibfnamefont{K.~B.}
  \bibnamefont{Whaley}}, \bibinfo{journal}{Physics Today}
  \textbf{\bibinfo{volume}{54}}, \bibinfo{pages}{31} (\bibinfo{year}{2001}).

\bibitem[{\citenamefont{Gough et~al.}(1985)\citenamefont{Gough, Mengel,
  Rowntree, and Scoles}}]{goyal85}
\bibinfo{author}{\bibfnamefont{T.~E.} \bibnamefont{Gough}},
  \bibinfo{author}{\bibfnamefont{M.}~\bibnamefont{Mengel}},
  \bibinfo{author}{\bibfnamefont{P.~A.} \bibnamefont{Rowntree}},
  \bibnamefont{and} \bibinfo{author}{\bibfnamefont{G.}~\bibnamefont{Scoles}},
  \bibinfo{journal}{J. Chem. Phys.} \textbf{\bibinfo{volume}{83}},
  \bibinfo{pages}{4958} (\bibinfo{year}{1985}).

\bibitem[{\citenamefont{Nauta and Miller}(2000)}]{nauta00}
\bibinfo{author}{\bibfnamefont{K.}~\bibnamefont{Nauta}} \bibnamefont{and}
  \bibinfo{author}{\bibfnamefont{R.~E.} \bibnamefont{Miller}},
  \bibinfo{journal}{Science} \textbf{\bibinfo{volume}{287}},
  \bibinfo{pages}{293} (\bibinfo{year}{2000}).

\bibitem[{\citenamefont{Bartelet et~al.}(1996)\citenamefont{Bartelet, Close,
  Federmann, Qaas, and Toennies}}]{bartelt96}
\bibinfo{author}{\bibfnamefont{A.}~\bibnamefont{Bartelet}},
  \bibinfo{author}{\bibfnamefont{J.~D.} \bibnamefont{Close}},
  \bibinfo{author}{\bibfnamefont{F.}~\bibnamefont{Federmann}},
  \bibinfo{author}{\bibfnamefont{N.}~\bibnamefont{Qaas}}, \bibnamefont{and}
  \bibinfo{author}{\bibfnamefont{J.~P.} \bibnamefont{Toennies}},
  \bibinfo{journal}{Phys. Rev. Lett.} \textbf{\bibinfo{volume}{77}},
  \bibinfo{pages}{3535} (\bibinfo{year}{1996}).

\bibitem[{\citenamefont{Federmann et~al.}(1999)\citenamefont{Federmann,
  Hoffman, Qaas, and Toennies}}]{federmann99}
\bibinfo{author}{\bibfnamefont{F.}~\bibnamefont{Federmann}},
  \bibinfo{author}{\bibfnamefont{K.}~\bibnamefont{Hoffman}},
  \bibinfo{author}{\bibfnamefont{N.}~\bibnamefont{Qaas}}, \bibnamefont{and}
  \bibinfo{author}{\bibfnamefont{J.~P.} \bibnamefont{Toennies}},
  \bibinfo{journal}{European Physical Journal D} \textbf{\bibinfo{volume}{9}},
  \bibinfo{pages}{11} (\bibinfo{year}{1999}).

\bibitem[{\citenamefont{Nauta et~al.}(2001)\citenamefont{Nauta, Moore, Stiles,
  and Miller}}]{nauta01}
\bibinfo{author}{\bibfnamefont{K.}~\bibnamefont{Nauta}},
  \bibinfo{author}{\bibfnamefont{D.~T.} \bibnamefont{Moore}},
  \bibinfo{author}{\bibfnamefont{P.~L.} \bibnamefont{Stiles}},
  \bibnamefont{and} \bibinfo{author}{\bibfnamefont{R.~E.}
  \bibnamefont{Miller}}, \bibinfo{journal}{Science}
  \textbf{\bibinfo{volume}{292}}, \bibinfo{pages}{481} (\bibinfo{year}{2001}).

\bibitem[{\citenamefont{Nauta and Miller}(1999)}]{nauta99}
\bibinfo{author}{\bibfnamefont{K.}~\bibnamefont{Nauta}} \bibnamefont{and}
  \bibinfo{author}{\bibfnamefont{R.~E.} \bibnamefont{Miller}},
  \bibinfo{journal}{Science} \textbf{\bibinfo{volume}{283}},
  \bibinfo{pages}{1895} (\bibinfo{year}{1999}).

\bibitem[{\citenamefont{Grebenev
  et~al.}(2000{\natexlab{a}})\citenamefont{Grebenev, Sartakov, Toennies, and
  Vilesov}}]{grebenev00a}
\bibinfo{author}{\bibfnamefont{S.}~\bibnamefont{Grebenev}},
  \bibinfo{author}{\bibfnamefont{B.}~\bibnamefont{Sartakov}},
  \bibinfo{author}{\bibfnamefont{J.~P.} \bibnamefont{Toennies}},
  \bibnamefont{and} \bibinfo{author}{\bibfnamefont{A.~F.}
  \bibnamefont{Vilesov}}, \bibinfo{journal}{Science}
  \textbf{\bibinfo{volume}{289}}, \bibinfo{pages}{1532}
  (\bibinfo{year}{2000}{\natexlab{a}}).

\bibitem[{\citenamefont{Kwon and Whaley}(2002)}]{kwon02}
\bibinfo{author}{\bibfnamefont{Y.}~\bibnamefont{Kwon}} \bibnamefont{and}
  \bibinfo{author}{\bibfnamefont{K.~B.} \bibnamefont{Whaley}},
  \bibinfo{journal}{Phys. Rev. Lett.} \textbf{\bibinfo{volume}{89}},
  \bibinfo{pages}{273401} (\bibinfo{year}{2002}).

\bibitem[{\citenamefont{Grebenev
  et~al.}(2001{\natexlab{a}})\citenamefont{Grebenev, Sartakov, Toennies, and
  Vilesov}}]{grebenev01a}
\bibinfo{author}{\bibfnamefont{S.}~\bibnamefont{Grebenev}},
  \bibinfo{author}{\bibfnamefont{B.~G.} \bibnamefont{Sartakov}},
  \bibinfo{author}{\bibfnamefont{J.~P.} \bibnamefont{Toennies}},
  \bibnamefont{and} \bibinfo{author}{\bibfnamefont{A.~F.}
  \bibnamefont{Vilesov}}, \bibinfo{journal}{J. Chem. Phys.}
  \textbf{\bibinfo{volume}{114}}, \bibinfo{pages}{617}
  (\bibinfo{year}{2001}{\natexlab{a}}).

\bibitem[{\citenamefont{Grebenev
  et~al.}(2001{\natexlab{b}})\citenamefont{Grebenev, Lugovoi, Sartakov,
  Toennies, and Vilesov}}]{grebenev01}
\bibinfo{author}{\bibfnamefont{S.}~\bibnamefont{Grebenev}},
  \bibinfo{author}{\bibfnamefont{E.}~\bibnamefont{Lugovoi}},
  \bibinfo{author}{\bibfnamefont{B.~G.} \bibnamefont{Sartakov}},
  \bibinfo{author}{\bibfnamefont{J.~P.} \bibnamefont{Toennies}},
  \bibnamefont{and} \bibinfo{author}{\bibfnamefont{A.~F.}
  \bibnamefont{Vilesov}}, \bibinfo{journal}{Faraday Discussions}
  \textbf{\bibinfo{volume}{118}}, \bibinfo{pages}{19}
  (\bibinfo{year}{2001}{\natexlab{b}}).

\bibitem[{\citenamefont{Tang and McKellar}(2002)}]{mckellar02}
\bibinfo{author}{\bibfnamefont{J.}~\bibnamefont{Tang}} \bibnamefont{and}
  \bibinfo{author}{\bibfnamefont{A.~R.~W.} \bibnamefont{McKellar}},
  \bibinfo{journal}{J. Chem. Phys.} \textbf{\bibinfo{volume}{116}},
  \bibinfo{pages}{646} (\bibinfo{year}{2002}).

\bibitem[{\citenamefont{Kwon and Whaley}(1999)}]{kwon99}
\bibinfo{author}{\bibfnamefont{Y.}~\bibnamefont{Kwon}} \bibnamefont{and}
  \bibinfo{author}{\bibfnamefont{K.~B.} \bibnamefont{Whaley}},
  \bibinfo{journal}{Phys. Rev. Lett.} \textbf{\bibinfo{volume}{83}},
  \bibinfo{pages}{4108} (\bibinfo{year}{1999}).

\bibitem[{\citenamefont{Paesani et~al.}(2001)\citenamefont{Paesani, Gianturco,
  and Whaley}}]{paesani01c}
\bibinfo{author}{\bibfnamefont{F.}~\bibnamefont{Paesani}},
  \bibinfo{author}{\bibfnamefont{F.~A.} \bibnamefont{Gianturco}},
  \bibnamefont{and} \bibinfo{author}{\bibfnamefont{K.~B.}
  \bibnamefont{Whaley}}, \bibinfo{journal}{J. Chem. Phys.}
  \textbf{\bibinfo{volume}{115}}, \bibinfo{pages}{10225}
  (\bibinfo{year}{2001}).

\bibitem[{\citenamefont{Higgins et~al.}(2003)\citenamefont{Higgins, Yu, and
  Klemperer}}]{higgins03}
\bibinfo{author}{\bibfnamefont{K.}~\bibnamefont{Higgins}},
  \bibinfo{author}{\bibfnamefont{Z.}~\bibnamefont{Yu}}, \bibnamefont{and}
  \bibinfo{author}{\bibfnamefont{W.~H.} \bibnamefont{Klemperer}}, p.
  \bibinfo{pages}{to be published} (\bibinfo{year}{2003}).

\bibitem[{\citenamefont{Zillich and Whaley}(2003)}]{zillich03a}
\bibinfo{author}{\bibfnamefont{R.}~\bibnamefont{Zillich}} \bibnamefont{and}
  \bibinfo{author}{\bibfnamefont{K.~B.} \bibnamefont{Whaley}},
  \bibinfo{journal}{Chem. Phys.} p. \bibinfo{pages}{in press}
  (\bibinfo{year}{2003}).

\bibitem[{\citenamefont{Kwon et~al.}(2000)\citenamefont{Kwon, Huang, Patel,
  Blume, and Whaley}}]{kwon00}
\bibinfo{author}{\bibfnamefont{Y.}~\bibnamefont{Kwon}},
  \bibinfo{author}{\bibfnamefont{P.}~\bibnamefont{Huang}},
  \bibinfo{author}{\bibfnamefont{M.~V.} \bibnamefont{Patel}},
  \bibinfo{author}{\bibfnamefont{D.}~\bibnamefont{Blume}}, \bibnamefont{and}
  \bibinfo{author}{\bibfnamefont{K.~B.} \bibnamefont{Whaley}},
  \bibinfo{journal}{J. Chem. Phys.} \textbf{\bibinfo{volume}{113}},
  \bibinfo{pages}{6469} (\bibinfo{year}{2000}).

\bibitem[{\citenamefont{McMahon et~al.}(1996)\citenamefont{McMahon, Barnett,
  and Whaley}}]{mcmahon96}
\bibinfo{author}{\bibfnamefont{M.~A.} \bibnamefont{McMahon}},
  \bibinfo{author}{\bibfnamefont{R.~N.} \bibnamefont{Barnett}},
  \bibnamefont{and} \bibinfo{author}{\bibfnamefont{K.~B.}
  \bibnamefont{Whaley}}, \bibinfo{journal}{J. Chem. Phys.}
  \textbf{\bibinfo{volume}{104}}, \bibinfo{pages}{5080} (\bibinfo{year}{1996}).

\bibitem[{\citenamefont{Huang et~al.}(2002)\citenamefont{Huang, Kwon, and
  Whaley}}]{huang02a}
\bibinfo{author}{\bibfnamefont{P.}~\bibnamefont{Huang}},
  \bibinfo{author}{\bibfnamefont{Y.}~\bibnamefont{Kwon}}, \bibnamefont{and}
  \bibinfo{author}{\bibfnamefont{K.~B.} \bibnamefont{Whaley}}, in
  \emph{\bibinfo{booktitle}{Quantum Fluids in Confinement}}, edited by
  \bibinfo{editor}{\bibfnamefont{E.}~\bibnamefont{Krotscheck}}
  \bibnamefont{and} \bibinfo{editor}{\bibfnamefont{J.}~\bibnamefont{Navarro}}
  (\bibinfo{publisher}{World Scientific}, \bibinfo{address}{Singapore},
  \bibinfo{year}{2002}), vol.~\bibinfo{volume}{4} of
  \emph{\bibinfo{series}{Advances in Quantum Many-Body Theories}},
  \eprint{physics/0204089}.

\bibitem[{\citenamefont{Patel et~al.}(2003)\citenamefont{Patel, Viel, Paesani,
  Huang, and Whaley}}]{patel02}
\bibinfo{author}{\bibfnamefont{M.~V.} \bibnamefont{Patel}},
  \bibinfo{author}{\bibfnamefont{A.}~\bibnamefont{Viel}},
  \bibinfo{author}{\bibfnamefont{F.}~\bibnamefont{Paesani}},
  \bibinfo{author}{\bibfnamefont{P.}~\bibnamefont{Huang}}, \bibnamefont{and}
  \bibinfo{author}{\bibfnamefont{K.~B.} \bibnamefont{Whaley}},
  \bibinfo{journal}{J. Chem. Phys.}  (\bibinfo{year}{2003}).

\bibitem[{\citenamefont{Zillich et~al.}(2003)\citenamefont{Zillich, Kwon, and
  Whaley}}]{zillich03b}
\bibinfo{author}{\bibfnamefont{R.}~\bibnamefont{Zillich}},
  \bibinfo{author}{\bibfnamefont{Y.}~\bibnamefont{Kwon}}, \bibnamefont{and}
  \bibinfo{author}{\bibfnamefont{K.~B.} \bibnamefont{Whaley}},
  \bibinfo{journal}{to be published}  (\bibinfo{year}{2003}).

\bibitem[{\citenamefont{van~den Bergh and Schouten}(1988)}]{bergh88}
\bibinfo{author}{\bibfnamefont{L.~C.} \bibnamefont{van~den Bergh}}
  \bibnamefont{and} \bibinfo{author}{\bibfnamefont{J.~A.}
  \bibnamefont{Schouten}}, \bibinfo{journal}{J. Chem. Phys.}
  \textbf{\bibinfo{volume}{89}}, \bibinfo{pages}{2336} (\bibinfo{year}{1988}).

\bibitem[{\citenamefont{Aziz et~al.}(1987)\citenamefont{Aziz, McCourt, and
  Wong}}]{aziz87}
\bibinfo{author}{\bibfnamefont{R.~A.} \bibnamefont{Aziz}},
  \bibinfo{author}{\bibfnamefont{F.~R.~W.} \bibnamefont{McCourt}},
  \bibnamefont{and} \bibinfo{author}{\bibfnamefont{C.~C.~K.}
  \bibnamefont{Wong}}, \bibinfo{journal}{Mol. Phys.}
  \textbf{\bibinfo{volume}{61}}, \bibinfo{pages}{1487} (\bibinfo{year}{1987}).

\bibitem[{\citenamefont{Higgins and Klemperer}(1999)}]{higgins99}
\bibinfo{author}{\bibfnamefont{K.}~\bibnamefont{Higgins}} \bibnamefont{and}
  \bibinfo{author}{\bibfnamefont{W.~H.} \bibnamefont{Klemperer}},
  \bibinfo{journal}{J. Chem. Phys.} \textbf{\bibinfo{volume}{110}},
  \bibinfo{pages}{1383} (\bibinfo{year}{1999}).

\bibitem[{\citenamefont{Ceperley}(1995)}]{ceperley95}
\bibinfo{author}{\bibfnamefont{D.~M.} \bibnamefont{Ceperley}},
  \bibinfo{journal}{Rev. Mod. Phys.} \textbf{\bibinfo{volume}{67}},
  \bibinfo{pages}{279} (\bibinfo{year}{1995}).

\bibitem[{\citenamefont{Kwon and Whaley}(2001)}]{kwon01}
\bibinfo{author}{\bibfnamefont{Y.}~\bibnamefont{Kwon}} \bibnamefont{and}
  \bibinfo{author}{\bibfnamefont{K.~B.} \bibnamefont{Whaley}},
  \bibinfo{journal}{J. Chem. Phys.} \textbf{\bibinfo{volume}{114}},
  \bibinfo{pages}{3163} (\bibinfo{year}{2001}).

\bibitem[{\citenamefont{Pollock and Ceperley}(1987)}]{pollock87}
\bibinfo{author}{\bibfnamefont{E.~L.} \bibnamefont{Pollock}} \bibnamefont{and}
  \bibinfo{author}{\bibfnamefont{D.~M.} \bibnamefont{Ceperley}},
  \bibinfo{journal}{Phys. Rev. B} \textbf{\bibinfo{volume}{36}},
  \bibinfo{pages}{8343} (\bibinfo{year}{1987}).

\bibitem[{\citenamefont{Ceperley and Pollock}(1992)}]{ceperley92}
\bibinfo{author}{\bibfnamefont{D.~M.} \bibnamefont{Ceperley}} \bibnamefont{and}
  \bibinfo{author}{\bibfnamefont{E.~L.} \bibnamefont{Pollock}}, in
  \emph{\bibinfo{booktitle}{Monte Carlo Methods in Theoretical Physics}},
  edited by \bibinfo{editor}{\bibfnamefont{S.}~\bibnamefont{Caracciolo}}
  \bibnamefont{and} \bibinfo{editor}{\bibfnamefont{A.}~\bibnamefont{Fabrocini}}
  (\bibinfo{publisher}{ETS Editrice}, \bibinfo{address}{Pisa, Italy},
  \bibinfo{year}{1992}).

\bibitem[{\citenamefont{Sindzingre et~al.}(1989)\citenamefont{Sindzingre,
  Klein, and Ceperley}}]{sindzingre89}
\bibinfo{author}{\bibfnamefont{P.}~\bibnamefont{Sindzingre}},
  \bibinfo{author}{\bibfnamefont{M.~L.} \bibnamefont{Klein}}, \bibnamefont{and}
  \bibinfo{author}{\bibfnamefont{D.~M.} \bibnamefont{Ceperley}},
  \bibinfo{journal}{Phys. Rev. Lett.} \textbf{\bibinfo{volume}{63}},
  \bibinfo{pages}{1601} (\bibinfo{year}{1989}).

\bibitem[{\citenamefont{Draeger and Ceperley}(2003)}]{draeger03}
\bibinfo{author}{\bibfnamefont{E.}~\bibnamefont{Draeger}} \bibnamefont{and}
  \bibinfo{author}{\bibfnamefont{D.}~\bibnamefont{Ceperley}},
  \bibinfo{journal}{Phys. Rev. Lett.} \textbf{\bibinfo{volume}{90}},
  \bibinfo{pages}{065301} (\bibinfo{year}{2003}).

\bibitem[{\citenamefont{BOUND}()}]{bound}
\bibinfo{author}{\bibnamefont{BOUND}}, \emph{\bibinfo{title}{computer code
  version 5, distributed by collaborative computational project no. 6 of the
  science and engineering research council (u.k.)}}.

\bibitem[{\citenamefont{Hunt et~al.}(1985)\citenamefont{Hunt, Foster, Johns,
  and McKellar}}]{hunt85}
\bibinfo{author}{\bibfnamefont{N.}~\bibnamefont{Hunt}},
  \bibinfo{author}{\bibfnamefont{S.~C.} \bibnamefont{Foster}},
  \bibinfo{author}{\bibfnamefont{J.~W.~C.} \bibnamefont{Johns}},
  \bibnamefont{and} \bibinfo{author}{\bibfnamefont{A.~R.~W.}
  \bibnamefont{McKellar}}, \bibinfo{journal}{J. Mol. Spectrosc.}
  \textbf{\bibinfo{volume}{111}}, \bibinfo{pages}{42} (\bibinfo{year}{1985}).

\bibitem[{\citenamefont{Hayman et~al.}(1989)\citenamefont{Hayman, Hodge,
  Howard, Muenter, and Dyke}}]{hayman89}
\bibinfo{author}{\bibfnamefont{G.~D.} \bibnamefont{Hayman}},
  \bibinfo{author}{\bibfnamefont{J.}~\bibnamefont{Hodge}},
  \bibinfo{author}{\bibfnamefont{B.~J.} \bibnamefont{Howard}},
  \bibinfo{author}{\bibfnamefont{J.~S.} \bibnamefont{Muenter}},
  \bibnamefont{and} \bibinfo{author}{\bibfnamefont{T.~R.} \bibnamefont{Dyke}},
  \bibinfo{journal}{J. Mol. Spectrosc.} \textbf{\bibinfo{volume}{133}},
  \bibinfo{pages}{423} (\bibinfo{year}{1989}).

\bibitem[{\citenamefont{Patel}(2001)}]{patel01a}
\bibinfo{author}{\bibfnamefont{M.}~\bibnamefont{Patel}},
  \bibinfo{journal}{Faraday Discussions} \textbf{\bibinfo{volume}{118}},
  \bibinfo{pages}{47} (\bibinfo{year}{2001}).

\bibitem[{\citenamefont{Grebenev et~al.}(2002)\citenamefont{Grebenev, Sartakov,
  Toennies, and Vilesov}}]{grebenev02}
\bibinfo{author}{\bibfnamefont{S.}~\bibnamefont{Grebenev}},
  \bibinfo{author}{\bibfnamefont{B.~G.} \bibnamefont{Sartakov}},
  \bibinfo{author}{\bibfnamefont{J.~P.} \bibnamefont{Toennies}},
  \bibnamefont{and} \bibinfo{author}{\bibfnamefont{A.~F.}
  \bibnamefont{Vilesov}}, \bibinfo{journal}{Phys. Rev. Lett.}
  \textbf{\bibinfo{volume}{89}}, \bibinfo{pages}{225301}
  (\bibinfo{year}{2002}).

\bibitem[{\citenamefont{Huang et~al.}(2003)\citenamefont{Huang, Sachse, and
  Whaley}}]{huang03}
\bibinfo{author}{\bibfnamefont{P.}~\bibnamefont{Huang}},
  \bibinfo{author}{\bibfnamefont{T.}~\bibnamefont{Sachse}}, \bibnamefont{and}
  \bibinfo{author}{\bibfnamefont{K.~B.} \bibnamefont{Whaley}},
  \bibinfo{journal}{to be published}  (\bibinfo{year}{2003}).

\bibitem[{\citenamefont{Gordy and Cook}(1984)}]{gordy}
\bibinfo{author}{\bibfnamefont{W.}~\bibnamefont{Gordy}} \bibnamefont{and}
  \bibinfo{author}{\bibfnamefont{R.~L.} \bibnamefont{Cook}},
  \emph{\bibinfo{title}{Microwave Molecular Spectra}}
  (\bibinfo{publisher}{Wiley Interscience}, \bibinfo{address}{New York},
  \bibinfo{year}{1984}).

\bibitem[{\citenamefont{Grebenev
  et~al.}(2000{\natexlab{b}})\citenamefont{Grebenev, Hartmann, Havenith,
  Sartakov, Toennies, and Vilesov}}]{grebenev00b}
\bibinfo{author}{\bibfnamefont{S.}~\bibnamefont{Grebenev}},
  \bibinfo{author}{\bibfnamefont{M.}~\bibnamefont{Hartmann}},
  \bibinfo{author}{\bibfnamefont{M.}~\bibnamefont{Havenith}},
  \bibinfo{author}{\bibfnamefont{B.}~\bibnamefont{Sartakov}},
  \bibinfo{author}{\bibfnamefont{J.~P.} \bibnamefont{Toennies}},
  \bibnamefont{and} \bibinfo{author}{\bibfnamefont{A.~F.}
  \bibnamefont{Vilesov}}, \bibinfo{journal}{J. Chem. Phys.}
  \textbf{\bibinfo{volume}{112}}, \bibinfo{pages}{4485}
  (\bibinfo{year}{2000}{\natexlab{b}}).

\bibitem[{\citenamefont{Quack and Suhm}(1991)}]{quack91}
\bibinfo{author}{\bibfnamefont{M.}~\bibnamefont{Quack}} \bibnamefont{and}
  \bibinfo{author}{\bibfnamefont{M.~A.} \bibnamefont{Suhm}},
  \bibinfo{journal}{J. Chem. Phys.} \textbf{\bibinfo{volume}{95}},
  \bibinfo{pages}{28} (\bibinfo{year}{1991}).

\end{thebibliography}


\end{document}